\documentclass[sigconf, natbib=false]{acmart}

\renewcommand\footnotetextcopyrightpermission[1]{} % removes footnote with conference information in first column
\setcopyright{none}
\AtBeginDocument{%
  \providecommand\BibTeX{{%
    \normalfont B\kern-0.5em{\scshape i\kern-0.25em b}\kern-0.8em\TeX}}}

\usepackage{xcolor}
\usepackage[T2A,T1]{fontenc}
\usepackage[utf8]{inputenc}
\usepackage[russian,english]{babel}
\usepackage{graphicx}
\usepackage{subfigure}
\usepackage{pgfplotstable}
\usepackage{booktabs}

\usepackage{tikz}

\newcommand\ChapterPrecis[2]{%
\begin{tikzpicture}[remember picture,overlay]
\node[anchor=north, draw=black, fill=yellow!20, inner sep=5pt, rounded corners, align=left, yshift=-#1] at (current page.north) 
{\parbox[t][2cm][l]{20cm}{\small #2}};
\end{tikzpicture}%
}

\providecommand{\ie}{\emph{i.e.,} }
\providecommand{\eg}{\emph{e.g.,} }

\usepackage{biblatex}
\acmConference[Preprint version]{25,10}{2021}

\begin{document}

\title{Uncovering the Dark Side of Telegram: Fakes, Clones,\\ Scams, and Conspiracy Movements}

\author{Massimo La Morgia}
\email{lamorgia@di.uniroma1.it}
\affiliation{%
  \institution{Sapienza University of Rome}
  \country{Italy}
}
\author{Alessandro Mei}
\email{mei@di.uniroma1.it}
\affiliation{%
  \institution{Sapienza University of Rome}
  \country{Italy}
}

\author{Alberto Maria Mongardini}
\email{mongardini@di.uniroma1.it}
\affiliation{%
  \institution{Sapienza University of Rome}
  \country{Italy}
}

\author{Jie Wu}
\email{jiewu@temple.edu}
\affiliation{%
  \institution{Temple University}
  \country{China}
}


\begin{abstract}
    Telegram is one of the most used instant messaging apps worldwide.
    Some of its success lies in providing high privacy protection and social network features like the channels---virtual rooms in which only the admins can post and broadcast messages to all its subscribers.
    However, these same features contributed to the emergence of borderline activities and, as is common with Online Social Networks, the heavy presence of fake accounts.
    Telegram started to address these issues by introducing the verified and scam marks for the channels. Unfortunately, the problem is far from being solved.
    
    In this work, we perform a large-scale analysis of Telegram by collecting 35,382 different channels and over 130,000,000 messages.
    We study the channels that Telegram marks as verified or scam, highlighting analogies and differences.
    Then, we move to the unmarked channels. Here, we find some of the infamous activities also present on privacy-preserving services of the Dark Web, such as carding, sharing of illegal adult and copyright protected content.
    In addition, we identify and analyze two other types of channels: the clones and the fakes. Clones are channels that publish the exact content of another channel to gain subscribers and promote services. Instead, fakes are channels that attempt to impersonate celebrities or well-known services. Fakes are hard to identify even by the most advanced users.
    To detect the fake channels automatically, we propose a machine learning model that is able to identify them with an accuracy of 86\%. Lastly, we study Sabmyk, a conspiracy theory that exploited fakes and clones to spread quickly on the platform reaching over 1,000,000 users.    
\end{abstract}




\maketitle

%%%%% Arxiv Reference
    \ChapterPrecis{0.3cm}{This paper is unpublished and represents an earlier version of published works. If you wish to cite it, please refer to the following instead:
    \begin{itemize}
        \item La Morgia, Massimo, et al. "It’s a Trap! detection and analysis of fake channels on telegram." 2023 IEEE International Conference on Web Services (ICWS). IEEE, 2023.
        \item La Morgia, Massimo, et al. "Pretending to be a VIP! Characterization and Detection of Fake and Clone Channels on Telegram." ACM Transactions on the Web (2024).
    \end{itemize}
    If you use the dataset, please cite:
    \begin{itemize}
        \item La Morgia, Massimo, et al. "TGDataset: Collecting and Exploring the Largest Telegram Channels Dataset." ACM SIGKDD Conference on Knowledge Discovery and Data Mining V.1 (KDD '25).
    \end{itemize}
}
%%%%

\section{Introduction}
%Nowadays, instant messaging apps are widespread. Among these, Telegram is becoming increasingly popular.
Telegram is likely the most controversial instant messaging platform. From one side, it has become extremely popular for its focus on user privacy, with tons of users that flocked from Whatsapp to Telegram in early 2021~\cite{fortune}, following the update of the privacy policy of Whatsapp. 
%One of its prominent features is user privacy: All messages stored within its servers are "heavily encrypted~\cite{telegramPrivacy}." 
On the other side, as is often the case with platforms that offer greater privacy protection, people abuse it for illegal purposes. Indeed, Telegram hit the news several times in the last years for the infamous activities run on the platforms.
%While it gives voice to dissidents in countries without freedom of speech~\cite{dissidents}, 
In Indonesia, terrorists used Telegram to promote radicalism and give instructions for carrying out attacks~\cites{bbcterrorist}. Neo-Nazi groups leverage Telegram to share their ideologies~\cite{telegramNeoNazi}.
Crypto investors coordinate large groups through Telegram channels to arrange market manipulations like pump and dump frauds~\cite{la2020pump}.
%also attracted the attention of terrorist organizations~\cite{cao2017dynamical} and groups with neo-Nazi ideologies~\cite{telegramNeoNazi}. 
%Cybercriminals, for example, use Telegram bots to collect user data stolen on phishing websites~\cite{gatherStolenData}. 
Moreover, to worsen everything, revenge porn~\cite{revengePorn} and channels with child pornography~\cite{chlidporn} content are not uncommon. 
All this shows that Telegram is a complex ecosystem with a largely unexplored dark side.

In this paper, we present TGDataset, a collection of more than $35,000$ channels. Analyzing the dataset, we find channels related to porn, carding, inciting violence, hacking, and white supremacism. 
We study two particular kinds of channels, the verified and the scam. The first are channels for whom Telegram verified that are official of a public figure, a creator, or a company. Instead, scam channels are those that users report to Telegram because the admins arrange frauds, impersonate public figures, or more, in general, channels that is better not to trust. We discover it is possible to reach scam channels following the flow of forwarded messages in less than 2 hops starting from a verified channel. Thus, unaware users can easily end up in those dangerous channels. Then, we deal with the widespread phenomenon of misleading accounts that affect online social networks and Telegram channels.
We define two different kinds of channels, the fakes and the clones.
%We study the phenomenon of phishing on Telegram, discovering that it is widespread and targets institutions, celebrities, companies, and often cryptocurrency exchange sites. 
%We define two kinds of channels: 
A fake channel is a channel pretending to be the official one of a celebrity or organization and posting messages different from those of the official one. A clone channel is a channel that mimics an official one publishing its exact content. 
Lastly, we carry out a thorough investigation on Sabmyk. Sabmyk is a new conspiracy theory that exploited fake and clones channels intensively to became popular.

Our main contributions are the following:
\begin{itemize}
    \item \textbf{TGDataset.} We build a new dataset made of $35,382$ channels. To the best of our knowledge, TGDataset is the first collection of Telegram channels that take a snapshot of the actual Telegram ecosystem instead of focusing on a particular topic. Moreover, we release our resource~\cite{dataset} publicly to help researchers in further investigations. 
    %\item \textbf{Telegram characterization}. We build a new dataset of 35,382 channels using seeds channels covering 18 different categories. 
    %We compare verified, scam, and standard channels and find that verified channels have more subscribers, more content, and their content is the most forwarded, while scam channels forward fewer messages. We build a directed graph that represents channels as nodes and forwarding between channels as edges. The graph reveals that scam channels are only a few hops away (on average two hops) from the verified ones. For more than half of verified channels, there is a path to reach all the scam channels.
    \item \textbf{Fake channels detection}. We define the phenomenon of fakes channels on Telegram, and we propose a machine learning model able to detect fake channels with an accuracy of $86\%$. With the proposed model, we detected $191$ allegedly fake accounts of which we could confirm $27$.
    %how that phishing is widespread on Telegram and we introduce a neural network model able to detect fake channels with an F1-score of $84\%$.
    \item \textbf{Clone channels analysis}. We find in our dataset $83$ clone channels. Analyzing them, we discover that the admins of clones exploit the popularity gained to promote cryptocurrencies, ideologies, or sell goods.
    %the behavior of clone channels and the methodologies they use to increase popularity. %Although it is easy to identify clone channels, they are very common as as way to gain many subscribers with low effort (on average, a clone channel has about 28,000 subscribers).
    \item \textbf{Sabmyk spread strategy}. We investigate Sabmyk. Analyzing our dataset as a graph, we identify the $98$ channels composing the Sabmyk network and analyze their strategy to reach a large audience quickly. 
    %find that this community exploits a network of popular channels, including fakes and clones, shares specific hot topics, and uses methodologies to forward content between the channels with the goal of maximizing the impact of conspiracy theories. 
\end{itemize}

\section{Telegram}
Telegram is a popular instant messaging platform that started in 2013, with more than 500 million active users by 2021~\cite{TelegramUsers}. On Telegram, users can share text messages, images, videos, audio, stickers, and files weighing up to 2~GB.
Aside from the standard one-to-one messaging, Telegram provides group chats and channels. 
Both have a unique username on the platform, a title, a description, and they can be private or public.
While groups allow many-to-many messaging (any member can write) and have a limit of 200,000 members, channels provide one-to-many communication (only admins can post content) and unlimited subscribers. Moreover, channels do not show info about the subscribers, except the total number. 
Although they serve different purposes, private chats, groups, and channels are not isolated but linked through message forwarding. This is a functionality that allows users and channels to forward content posted in a chat to a different user, group, or channel showing the author of the original message.
In particular, Telegram channels are an effective solution for spreading information to a large pool of people.
Indeed, several institutional public figures and companies opened an official Telegram channel to broadcast announcements and news~\cite{covidTelegram}. Likewise, start to pop up on the platform also channels aiming to impersonate these kinds of channels or that leverage Telegram channels and groups to sell fake products or services.  
To face this phenomenon, Telegram introduced the \textit{verified} and the \textit{scam} marks. Channels, groups, and bots can achieve the verified mark proving to Telegram that the profile has the verified status at least on two social media platforms (\eg TikTok, Facebook, Twitter, Instagram)~\cite{howVeryfyTelegramCh}.
Instead, Telegram flags a channel or a group as a scam if several users reported it for fraud~\cite{telegramScam}. 

\section{The dataset}
\label{sec:dataset}
In this section, we describe the approach we used to build our dataset, the TGDataset. Then, we explore its main characteristics, focusing primarily on verified, scam, and standard channels. Finally, we investigate the topics covered by the retrieved channels.

\subsection{Building the dataset}
\label{sec:bulding_dataset}
Existing Telegram datasets are designed for specific studies. Thus, they contain only channels related to a particular topic, language, or country. For instance, ~\cite{hashemi2019telegram} contains exclusively Iranian channels; while~\cite{hoseini2021globalization} only has channels related to QAnon. Finally, the PushShift dataset~\cite{baumgartner2020pushshift} focuses on right-wing extremist politics or cryptocurrencies related channels.
Conversely, our work aims to study the Telegram ecosystem broadly to understand how potential malicious actors behave and the main topics discussed on the platforms. Thus, we need a dataset representing an actual snapshot of Telegram covering many popular and connected channels. For these reasons, we build the TGDataset.

To explore Telegram, and in particular the most popular and connected channels, we start from seed channels covering different topics and expand the dataset by adding, for every forwarded message in the seed channels, the original channel of the message, as previously done in~\cite{baumgartner2020pushshift}. 
To select the seed channels, we leverage Tgstat~\cite{tgstat}, a popular service that indexes more than $150,000$ Telegram channels and collects statistics about them.
Tgstat reports for each channel information such as the number of subscribers, category (topic), growth percentage, and language. Among the other statistics, Tgstat reports the rank of the top $100$ channels by the number of users.
From this rank we retrieve all the categories these channels belong to, finding the following $18$ categories: \textit{Sales}, \textit{Humor and entertainment}, \textit{News and Mass media}, \textit{Video \& Movies}, \textit{Business \& Startups}, \textit{Cryptocurrencies}, \textit{Politics}, \textit{Technologies}, \textit{Sport}, \textit{Marketing}, \textit{Economics}, \textit{Games}, \textit{Religion}, \textit{Software and Applications}, \textit{Lifehacks}, \textit{Fashion \& Beauty}, \textit{Medicine}, \textit{Psychology}, and \textit{Adults}.
Then, we select as seeds the $10$ most popular channels by the number of subscribers from each category.

Overall, we obtain a total of $180$ seed channels. 
From each seed channel, we download the last $10,000$ messages through the Telethon APIs~\cite{telethonAPI}, an open-source Python wrapper of the official Telegram APIs. Although a channel can contain more than $10,000$ messages, we decide not to download more than that. Indeed, we empirically notice that $10,000$ messages are a fair trade-off to have a good representation of the channel's content and not abuse the Telegram's API. After downloading the data, we parse the messages to discover new channels analyzing the forwarded messages.
Finally, to further expand the TGDataset, we use the newly discovered channels as new seeds and iterate the above-described procedure. After three iterations, more than 50\% of seeds do not contribute any more since all theirs forwarding flows are completely explored.

From each channel, we store the following information: The title, the description, the username, the ID, the creation date, the number of subscribers and if it is marked as a scam or verified. 
Whereas for the messages, we store the author, the timestamp, and, in case of forwarded messages, the original author, the original posting date, and which is the original channel.
Finally, we store the content of the text-based messages, while just the title and the file format of the media-based messages.

\subsection{Dataset overview}
Data collection ended on 15 March 2021.
With the described approach, we found $36,673$ channels, 1,205 of which were not longer accessible and $86$ deleted.
%of which $35,382$ ($96.5\%$) were reachable, and the others were not accessible as they were private ($1,205$) or deleted ($86$).
Overall, the TGDataset is 121 GB in size, contains $132,804,557$ messages and $35,382$ different channels. Among the  channels, $191$ ($0.5\%$) are verified channels and $26$ ($0.07\%$) are scam channels. Manually investigating the scam channels, we found that they are mainly related to trading, cryptocurrencies, and fake accounts of political figures (\eg Donald Trump, Mike Pompeo, and Ivanka Trump). For the sake of clarity, we will refer to the $35,165$ channels that are neither scam nor verified as \textit{standard channels}.
In the following section, we analyze these three kinds of channels separately to highlight differences in behavior. 

\begin{figure}
  \subfigure[Subscribers]{%
  \includegraphics[width=.23\textwidth]{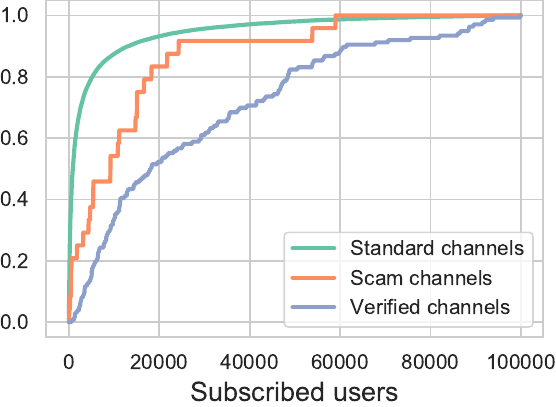}
  \label{fig:cdf_subscribers}}
  \subfigure[Forwarded messages]{%
  \includegraphics[width=.23\textwidth]{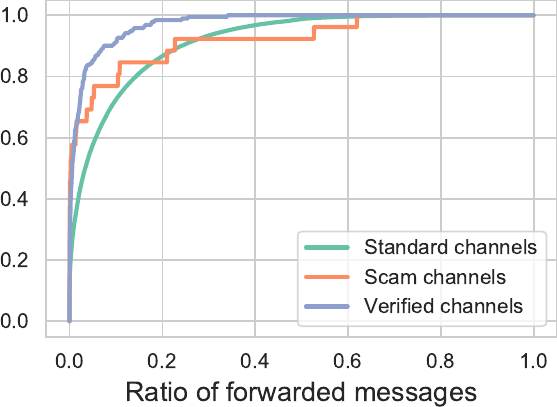}
  \label{fig:cdf_forwarded_messages}}
  \caption{\ref{fig:cdf_subscribers} CDF of the number of subscriber for scam, verified and standard channels.~\ref{fig:cdf_forwarded_messages} CDF of the ratio of forwarded messages for scam, verified an standard channels.} 
  \label{fig:cdf_dataset}
\end{figure}

\subsection{Subscribers and Messages}

As shown in Fig.~\ref{fig:cdf_subscribers}, the most popular of the three kinds of channels are the verified ones. Indeed, these channels represent celebrities or services, and hence it is very likely that they have a large base of subscribers.
On average, a verified channel has $181,887$ subscribers, with the most popular channel in this category (\textit{Telegram News}, the official channel of Telegram) having $5,478,200$ subscribers, while the smaller one $603$ (\textit{Russian MFA}, the official channel for Russian Foreign Ministry). Scam channels are the runner-up on this statistic, with an average number of subscribers of $31,420$. This result is not completely surprising. Indeed, the $7$ most popular scam channels attempt to impersonate popular people, and careless users could not recognize that these channels are not official. For instance, the most popular scam channel, with $364,663$ subscribers, aims to represent \textit{"Donald J Trump"} (https://t.me/trumps).
Lastly, we have the standard channels. Among them, the largest channel is \textit{HINDI HD MOVIES KGF LATEST}, a channel with $6,585,602$ subscribers for downloading copyrighted films. Nevertheless, standard channels usually have far fewer users, indeed $60\%$ of these channels (the beginning of the knee in Fig.~\ref{fig:cdf_subscribers}) have less than $1,541$ subscribers.

\begin{figure}
  \subfigure[Text-based messages]{%
  \includegraphics[width=.23\textwidth]{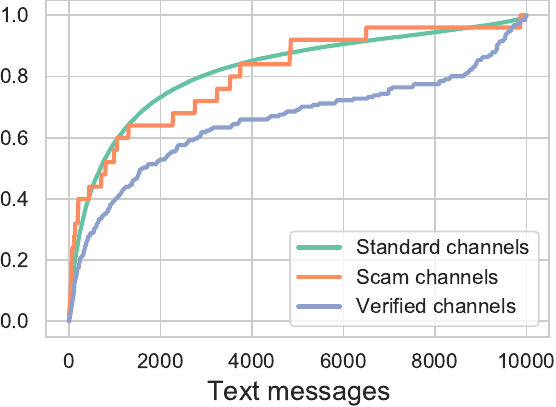}
  \label{fig:cdf_text_messages}}
  \subfigure[Media-based messages.]{%
  \includegraphics[width=.23\textwidth]{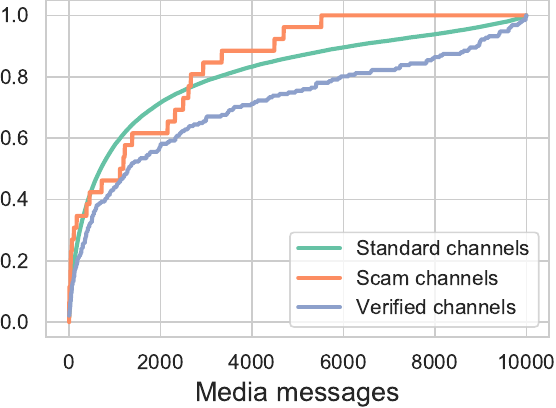}
  \label{fig:cdf_media_messages}}
  \caption{\ref{fig:cdf_text_messages} and \ref{fig:cdf_media_messages} display respectively the CDFs of the number of text-based and media-based messages posted by the 3 kind of channels.} 
  \label{fig:cdf_dataset_2}
\end{figure}

Then, we analyze the number of text-based and media-based messages shared by the three categories of channels.
Fig.~\ref{fig:cdf_text_messages} and Fig.~\ref{fig:cdf_media_messages} reveal that verified channels tend to share more messages in general, both text-based or media-based. On average, verified channels post $3,436.75$ text messages and $2,875.83$ media content, while scam and standard channels post $2,216.07$ and $1,835.77$ text messages and $1,539.57$ and $1,890.64$ media, respectively.
As displayed in Fig.~\ref{fig:cdf_forwarded_messages}, verified and
scam channels forward fewer messages than standard ones: on average, verified and scam channels forward $2.6\%$ and $7.5\%$ of their messages, respectively, while $8.3\%$ of the messages posted in standard channels are forwarded. 
Of these scam channels, the one that forwards more messages ($649$ messages from $94$ different channels) is \textit{Hype Royale}, a hacking channel for a mobile game.

\subsection{The Graph of the Dataset}
\label{sec:graph_analysis}

The dataset can be represented as a directed graph $G=(V, E)$ in which nodes in $V$ are the channels and edge $u \rightarrow v$ in $E$ represents the presence in channel~$u$ of a message originally posted in $v$ and forwarded to~$u$ by the admin of channel~$u$. Since the users of channel~$u$ can navigate the forwarded message and land on channel~$v$, the edge represents in an natural way the possible flow through channels of a users following forwarded messages.

The resulting graph has 7,551 strongly connected components, a giant component of 27,672 nodes (78\% of the channels), 139 components of two or more nodes, and 7,412 isolated nodes. 
The giant component includes 54 seed channels, while 50 seed channels are actually isolated nodes.
The main Tgstat categories to which these 50 channels belong are Fashion \& Beauty (7), Software and Application (6), News and mass media (5), and Humor and entertainment (5). The channel that posts forwarded messages from the largest set of other channels (\ie the node with largest out-degree), is \textit{Rekt plebs}, an entertainment channel that jokes about the losses of unwary cryptocurrency investors, which posts forwarded messages from 3,308 other channels, covering 9\% of the channels in the dataset. Instead, the channel whose messages are posted as forwarded messages to the largest set of other channels (\ie the node with largest in-degree), is a Russian news channel (\textit{\foreignlanguage{russian}{Раньше всех. Ну почти.}}) with in-degree 3,371.

One intriguing detail to investigate is whether and how seed channels are connected to scam channels. It is interesting since they are the most popular and, thanks to their status, could be considered trusted by the users. Looking at the distance (shortest path) between seed and scam channels, we notice that all seed channels are very close to the scam channels. Twelve of the scam channels can be reached with two hops and another dozen with three hops from a seed channel. One of the seed channels, \textit{Donald J. Trump}, is even itself a scam channel. In the case of verified channels, we find a similar situation. Indeed, 22 out of 26 scam channels can be reached with only two or three hops from verified ones. Furthermore, 116 verified channels ($60.7\%$ of the total number) are connected to all 26 scam channels.
These results show that it can be really easy that a user of a verified channel navigates to a scam channel.

Scam channels, between themselves, are almost isolated. Sixteen of the 26 scam channels are not connected to any other scam channel. The remaining ten are at an average distance of 5 hops from the closest scam channel, and one of the scam channels is 11 hops away from the nearest. Thus, it is likely that there is no collaboration between the scam channels in our dataset. Verified channels are the opposite. 120 of them ($62.8\%$) are less than 3 hops away from the closest verified channel, with 58 channels at only 1 hop. Nonetheless, we also find 62 verified channels disconnected from the other ones.

%there is a  between verified channels,  
%they are either isolated from each other or directly in contact (hop distance equal to 1). 
%there are not strong connection
%how scam channels are connected among them, we discover 
%Instead, for most scam channels ($16$ out of $26$), there is no path connecting them. This seems reasonable since scam channels have no interest in redirecting users to other scam channels (unless the same person runs them). 
%This is quite plausible since a verified channel can forward messages of other like-minded verified channels but, on the other hand, does not want to refer its subscribers to other competing channels as well.

\begin{figure}
  \subfigure[Page Rank values]{%
  \includegraphics[width=.23\textwidth]{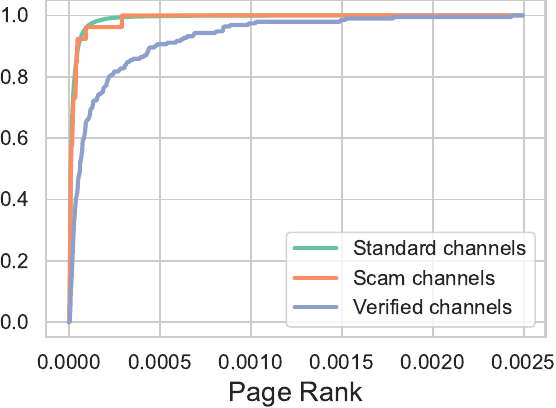}
  \label{fig:page_rank}}
  \subfigure[Copied messages]{%
  \includegraphics[width=.23\textwidth]{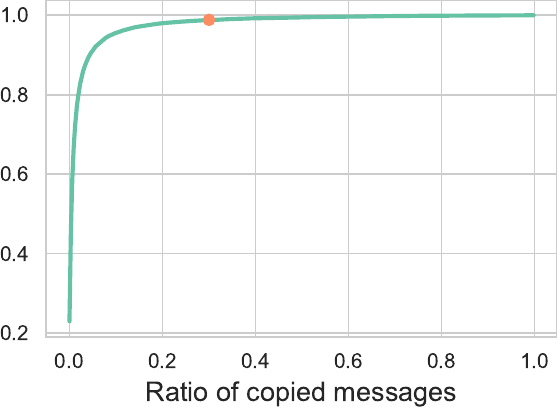}
  \label{fig:cdf_copied_messages}}
  \caption{\ref{fig:page_rank} CDF of PageRank values of scam, verified and standard channels.~\ref{fig:cdf_copied_messages} CDF of the ratio of copied messages.} 
  \label{fig:cdf_dataset_3}
\end{figure}

Another interesting aspect is related to which are the most influential channels. These are the nodes that spread the information more frequently and faster~\cite{kempe2005influential}. One of the most popular approaches to identify the influential nodes is to use centrality metrics like PageRank~\cite{brin1998anatomy,chen2013identifying}. The idea is to define as the most relevant nodes those that have the highest PageRank. 
%\begin{figure}
%  \includegraphics[width=.44\textwidth]{images/page_r.pdf}
%  \caption{CDF of PageRank values of scam, verified and standard channels.}
%  \label{fig:page_rank}
%\end{figure}
Fig.~\ref{fig:page_rank} shows the CDF of the Page Rank values for verified, scam, and standard channels. Verified channels have a higher Page Rank value. In contrast, standard and scam channels have a similar distribution.
This shows that verified channels are the influential nodes within the graph and the main engine of information dissemination within Telegram. 
Interestingly, these results show that the Page Rank could be a very relevant feature to identify verified channels.

\subsection{Discovering Topics}
\label{sec:topic_modeling}

In this subsection, we investigate the topics covered by the channels in our TGDataset using Topic Modeling~\cite{hofmann2001unsupervised}.
This is a data mining tool that allows finding a brief description of the topic addressed by the messages of a channel.  
For this analysis, we consider only channels that post material in English.
We first pre-process the messages normalizing and polishing them. Then, to detect the languages of the channels, we leverage LangDetect~\cite{nakatani2010langdetect}, a language detection library implemented by Google with precision over $99\%$ for 53 languages. 
In this way, we find 7,101 ($20\%$ of our dataset) channels that we can use as input to the topic modeling.

To discover the latent topics addressed within the channels, we use the Latent Dirichlet Allocation (LDA)~\cite{blei2003latent} as Topic Modeling algorithm. LDA needs as input the number of topics, so we used the UMass measure~\cite{mimno2011optimizing} to select the optimal one.
LDA relies on the idea that documents are generated by a particular probabilistic model, according to which each document is composed of a mixture of a small number of topics, and each word belongs to one of them. 
UMass is an intrinsic measure of Topic Coherence~\cite{stevens2012exploring}. It computes the log-likelihood that two words that represent the topic occur in the same documents. 
In particular, the higher the coherence of the words representing topics, the closer to 0 the value of UMass. In our case, we calculate the best UMass value reached by selecting the number of topics from 10 to 25. The best is obtained with 14 topics, with a UMass value of -0.97. Tab.~\ref{tab:LDA_topics} in Appendix shows the inferred topics and top 10 keywords for each of them.  

Having identified the topics, we group the channels according to the topic.
We use the values obtained from LDA as input features to K-means~\cite{hartigan1979ak}, a classical clustering algorithm based on partitioning.
\begin{figure}
  \includegraphics[width=.44\textwidth]{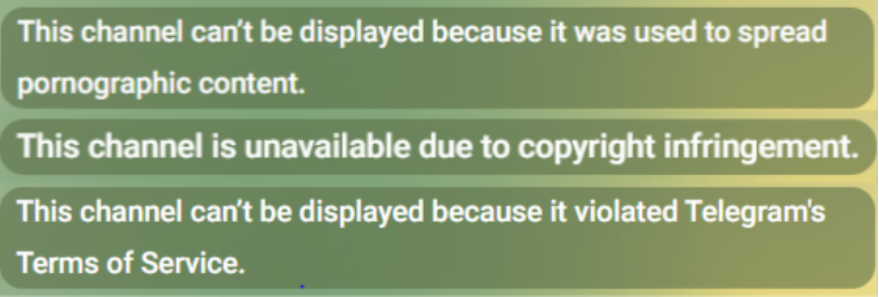}
  \caption{Telegram service messages. %related to violation of terms of service.
  }
  \label{fig:violeted_terms}
\end{figure}
Tab.~\ref{tab:english_topics} reports the topics discovered in the TGDataset and the number of channels associated to each topic. As we can see, the emerged topics are quite different from those covered by the seed channels (see Sec.~\ref{sec:bulding_dataset}).
Indeed Sport, Marketing, Humor and entertainment, Sales, Business Startups, Medicine, Psychology, and Fashion \& Beauty disappear from our dataset, while some other interesting topics come up.
One of these is related to carding.
Carding is the practice of selling full details of stolen credit cards or selling prepaid cards or other goods purchased with them. 
Similarly to what happens in dark web forums~\cite{kigerl2020behind}, carders (the people who own the stolen credit cards) use Telegram channels to place gift cards or goods for sale. In the TGDataset, we find 74 channels ($1\%$ of the English part of the dataset) that offer this service. Another unusual cluster of channels is about violated terms. They are 110 ($1.5\%$) channels whose messages have been obscured by Telegram because they incited violence, published illegal pornographic content, or shared content protected by copyright. In this case, Telegram replaced some or all of the channel's messages with text explaining the reasons for obfuscation, as those reported in Fig.~\ref{fig:violeted_terms}. Interestingly, the channel itself and the metadata (\eg the posting date) related to the original messages are still available. 
Despite the commitment of Telegram in obfuscating these channels, we still find public channels within our dataset promoting the spread of neo-Nazi ideologies or call for violence (\eg \textit{White Aryan Woman}, \textit{Feuerkrieg Division **OFFICIAL**}).
Therefore, the problem is still far from being solved.
The channels mentioned above belong to the group of channels that we identify as \textit{Religion and supremacism}, the largest group (3,352 channels) in the TGDataset.
In this group, there are channels strictly related to religion and channels that praise the supremacy of the white race. Intrigued by this surprising mixture of topics discovered in our dataset, we dig more on these channels, reading the shared content. We discover that the supremacist channels in this group use many references to the Christian religion or utilize religion itself as a motivation for their ideology. Thus, it is likely that the mixed tones used in this kind of channels lead our topic modeling approach to build this peculiar topic. Finally, we find that the other topics are more aligned with the ones covered by the seed channel of the TGDataset.

\begin{table}
    \centering
    \small
    \caption{Discovered topics and number of channels}
    \label{tab:english_topics}
    \begin{tabular}{l|c|c|c}
    \toprule
        Topic & \# channels & \# scam & \# verified\\ 
        \midrule
        Religion and supremacism & 3352 (47.20\%) & 1 & 5\\
        News &  1244 (17.52\%) & 8 & 30\\
        India news/career & 651 (9.17\%) & 0 & 4\\
        Adult Content      & 564 (7.94\%) & 0 & 1\\ 
        Games hacking & 269 (3.79\%) & 1  & 0\\
        Free music/movie & 220 (3.10\%) & 0 & 2 \\
        Software & 208 (2.93\%) & 0 & 9 \\
        Cryptocurrencies & 206 (2.90\%) & 0 & 0 \\
        Violated terms/pornographic & 110 (1.55\%) & 0 & 0\\
        Hacking & 103 (1.45\%) & 0 & 0 \\
        Carding & 74 (1.04\%) & 1  & 0 \\
        Telephone modding & 40 (0.56\%) & 0 & 0 \\
        Trump supporters & 39 (0.55\%) & 0 & 6 \\
        Games discussion & 21 (0.30\%) & 0 & 1 \\
        \bottomrule
    \end{tabular}
\end{table}

\section{Clone channels}
\label{sec:clones}
A curious aspect of Telegram is the presence of sets of two or more channels that post identical messages.
Clearly, the actual creator of the content is only one in the set, and we refer to it as the original channel. Instead, we call \textit{clone channels} the channels that publish the exact content of the original one.
To understand the reasons behind the creation of a clone channel and how common this phenomenon is, we examine the TGDataset.  

To find the clone channels, we compare the messages of each channel with those of all other channels. To reduce the search space significantly, we compare only channels written in the same language.
To avoid messages that could be coincidentally identical, we only take into considerations messages longer than 5 words, and we do not consider forwarded messages or messages about the violated terms, such as the ones in Fig.~\ref{fig:violeted_terms}.
Finally, we analyze the distribution of copied messages in our dataset (Fig~\ref{fig:cdf_copied_messages}). As we can see, more than $95\%$ of channels have less than $10\%$ identical messages in common with other channels. To find the clones, we restrictively select the tail of the CDF (the orange dot in the figure) that represents the channels with $30\%$ or more identical messages with another channel. We consider the channel $B$ a clone of the original channel $A$ if, for each common message, the one of $B$ has a publication date later than that of $A$.

With this approach, we find 83 clone channels. In particular, we find 37 written in English ($41.5\%$), 20 in Russian ($22.4\%$), and the others in Bulgarian, Farsi, German, Estonian, Hindi, Indonesian, Marathi, and Arabic.  Manually investigating the English channels, we find that the target of a clone is often the official channel of a celebrity or service. 
In particular, 5 clones have a different name with respect to the original channels but they post all the messages of the original ones. Moreover, they interleave the original messages with links to an external platform to buy goods (\eg books, microwaves) or links to join other channels. For instance, we find a clone of a cryptocurrency-related channel that promotes another channel that arranges pump and dumps operations~\cites{la2020pump}. 
5 channels clone the official channel of a celebrity and have very similar name of the original. In these clones, we find additional messages with controversial political content such as anti-vaccine campaigns. 
Then, we find a group of 10 channels cloning channels of politicians close to Donald J.\ Trump or Republican news channels. In this case, all the messages not taken from the original channels promote a new cryptocurrency called \textit{Trump coin}.

There are also 2 perfect clones with the same content, title, description, and profile image of another channel. These 2 channels copied the original channel for weeks and then started to post messages about a new conspiracy theory called Sabmyk (see Sec.~\ref{sec:sabmyk}).

Interestingly, we find 4 clones that, as the original channels, post books protected by copyright. We believe that the admin of the clones is the same admin of the original channels and uses the clones as a backup of the material shared. If this is the case, this technique appears to be effective. Indeed, checking the original channels a month later we downloaded the data, we found that Telegram removed the content of the original channels while the clones continued its activity.
Finally, for the other clones, we notice nothing suspicious other than being clones. However, it is crucial to remark that they are the clones with fewer subscribers (less than 1,000).
Thus, they could not have awakened yet, or the admin stopped his cloning activity, as we found in one case. It is clear that looking at their behavior, the goal of clone channels is to take advantage of the popularity and content generated by the original channel to gain subscribers and promote other services. 
This strategy is very effective. Indeed, the average number of subscribers of the clone channels is 28,491.06. The larger clone channel is one of the perfect clones, with over 100,000 subscribers. It is not surprising since, in this case, the clone and the official channel are virtually indistinguishable without knowing the channel's username. 

%Although we find less than $100$ clone channels, we believe this phenomenon is widely spread on Telegram, and further investigation is needed.

\section{Fake channels}

%In addition, phishing often occurs within cryptocurrency groups where a member acts as an admin and asks for sensitive information from other members. This phenomenon is not limited to cryptocurrency groups but also cryptocurrency exchange sites. Since there were no related official channels, some fake channels presented themselves as the official ones of Coinbase~\cite{coinbaseFake} and Kraken~\cite{fakeKraken}, two exchange sites.  

\label{sec:fake_detection}
Just like what happens with fake accounts in Online Social Networks~\cite{cao2012aiding, xiao2015detecting}, fake channels are widespread in Telegram.
A fake channel, as a fake account, attempts to impersonate an important service or person. 
A characteristic of fake channels is that the title is the exact name of the target or a slight variation of it (\eg presence of emoji in the title). Indeed, they attempt to qualify themselves as officials using words like official, real, and verified or adding the verified mark on the profile image.  
Fake channels are different from clone channels since they do not replicate the messages of the original channel.
%This section first analyzes how hard it is to distinguish fake channels from the officials with a user study \amnote{da modificare se non mettiamo lo user study}. 
In this section, firstly, we present our machine learning model to detect fake channels. Then, we apply our detector to the TGDataset.

\subsection{The Fake Channels dataset}
Training a machine learning model able to detect fake channels requires a ground-truth, a good amount of channels for whom we are sure about their status of official or fake.
Thus we create the Fake Channels dataset.
To build it, we use the following approach: We firstly leverage the \textit{telemeter.io}~\cite{telemeterIO} services to retrieve a list of verified channels. Then, for each verified channel found, we look for fake channels claiming to be the official ones.  
At the end of the process, the Fake Channels dataset consists of 342 different channels, 184 of which official and 158 fake. Of course, we discarded from the Fake Channels dataset clones or channels of the TGDataset, so that we can use the dataset as the training set.

\subsection{Features and classifier}
%The proposed classifier relies on the idea that it is possible to distinguish fake channels from official channels on message writing, description, title, reference to other channels, and the number of posted links. 
To build a classifier that detects fake channels, we leverage what we learned in Sec.~\ref{sec:dataset} along with other features that capture differences between official and fake channels in the writing style, references to other channels, and time of activity. We tried several sets of features and classifiers to build our model. In the following, we describe the configuration that achieved the best performances.    %analyzing scam and verified channels and the idea that fake channels differs form official  . Moreover we   
%use several features that leverage the writing style, the temporal aspect, and the external interaction.

\textbf{Writing style features:} average message length, average number of emojis per message, average number of non-alphanumeric characters per message, number of non-alphanumeric characters in the title and description, and average number of non-alphanumeric characters in the channel's title.

\textbf{Temporal features}: number of text messages published in the last 3, 6, 9 months and average posting time between two consecutive messages.

\textbf{External interaction features}: number of forwarded messages, standard deviation of the number of source channels for the forwarded messages, number of shared links, and number of duplicate messages containing at least one link.

We use these features to train a Multilayer Perceptron (MLP)~\cite{gardner1998artificial} of $4$ linear layers with Rectified Linear Unit function (ReLU)~\cite{hara2015analysis} as the activation function, Adam optimization algorithm~\cite{zhang2018improved} as the optimizer, and binary cross-entropy (BCE)~\cite{mannor2005cross} as the loss function. We train the model for 50 epochs and evaluate its performance through 5-fold cross-validation~\cite{anguita2012k}, achieving an accuracy of \textbf{$86\%$} and a weighed F1-score of {$85.7\%$}. 
To further assess our model, we run the detector on the verified and scam channels we have in our dataset. 
To experiment, we select all the 191 channels and the scams channel that attempt to impersonate users or services, accounting for $17$ channels.
Here the model detects as official $167$ out of $191$  channels and $12$ fakes out of $17$. Thus, in this experiment, the model achieves a global weighed F1-score of $88\%$ and an accuracy of $86\%$.

\begin{figure}
  \includegraphics[width=.44\textwidth]{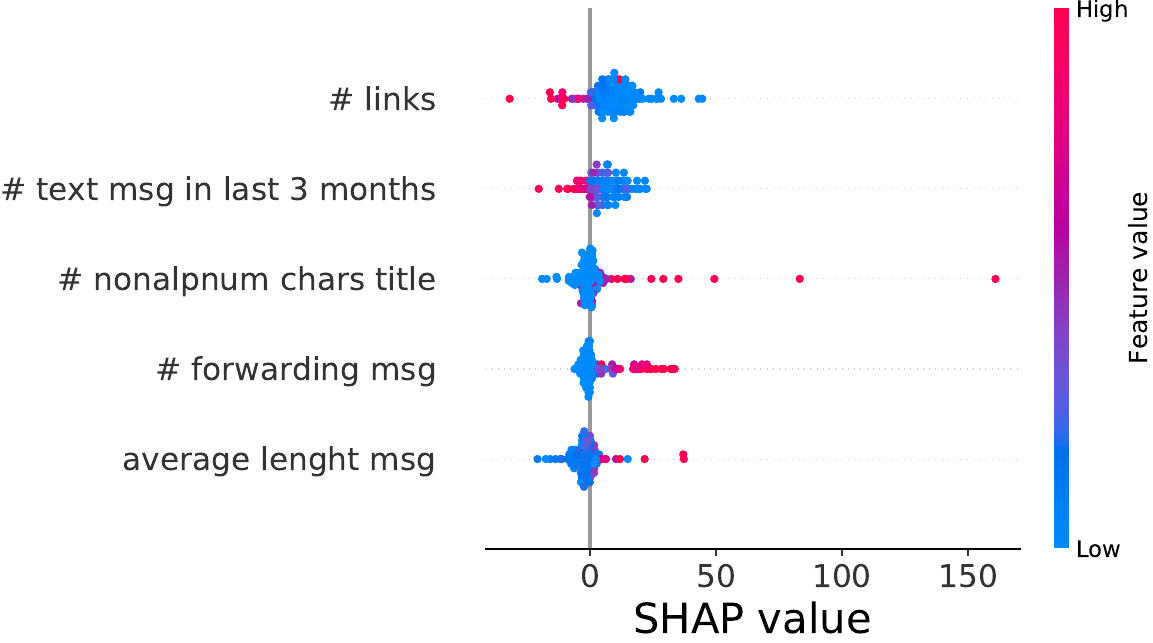}
  \caption{SHAP values of the 5 features that contribute most to model prediction.}
  \label{fig:shap_values}
\end{figure}

\subsection{Features analysis} 
To understand which features contribute more to obtain the excellent performance of our model, we use the Shapley Additive Explanations (SHAP) value~\cite{lundberg2017unified}. 
It determines the contribution of each feature based on game theory principles and local explanations. Fig.~\ref{fig:shap_values} shows the SHAP values of the 5 features that contribute more to the predictions of the model. 
According to the SHAP value, the 3 most significant features are the number of links posted within a channel, the number of text messages posted in the last 3 months, and the number of non-alphanumeric characters in the title. Interestingly, a high number of links suggests to the model that the channel is an official one. Indeed, analyzing the Fake Channels dataset, we find that the official channels tend to post many more social media links (on average 2006.24) than fake channels (on average 682.74). 
Moreover, a large number of posts published in the last 3 months inclines the model to consider a channel as an official. The cause could be that some fake channels, unlike the official ones, tend to have a short life of activity.
Instead, non-alphanumeric characters in the title lead the model to flag a channel as fake since several fake channels use emojis (especially the one similar to the verified channel symbol) in their title to attract users.  

Surprisingly, during our study, we notice that using as a feature the number of subscribers produces a negative effect on the model's performance. 
The reason for this is that the number of subscribers highly depends on the popularity of the target channel. In fact, several fake channels have more subscribers than official channels representing niche services or not so famous people. In Sec.~\ref{sec:graph_analysis} we saw that verified channels have a very high Page Rank value with respect to other kinds of channels. So, Page Rank could be a powerful feature to improve the model. However, to use the Page Rank, it is required to know the graph of the Telegram channel's, which is not even possible. For this reason, we do not use the Page Rank as a feature in our experiment.

\subsection{Discovering fake channels}
After validating our classifier, we leverage it to detect fake channels on the TGDataset.
For this task, we consider only English channels. 
We ignore channels dealing with carding, hacking, and video games, as they do not have websites from which it is possible to verify their identity.  
Hence, we collect the channels that have in their title, description or username the words \textit{real}, \textit{official} or \textit{verified}. Indeed, as we said, several fake channels use these words to persuade users that they are official. To further expand the dataset, we consider all the channels that have a similar name (edit distance less than 3) to one of the verified channels. 
In the end, we collected a set of 502 channels.
The classifier returns as fake 198 channels out of 502.
Since we do not have a ground truth for this set of channels, we check all of them manually to assess the results.
In particular, we consider a channel: 
\begin{itemize}
    \item \textbf{Official}: if Telegram marked it as verified or there exists an official source (\eg Website, Facebook, Instagram, Twitter) of the person/service indicating the Telegram channel as the official one. 
    \item \textbf{Fake}: if there is another channel that we consider official with the same name or an official source states that there is no official Telegram channel.
    \item \textbf{Allegedly fake/official}: if our classifier detects the channel as fake/official, but there is no evidence of their status. In particular, we have no channel with the same or a similar name that we consider official and the related official web pages or social media pages do not mention any Telegram channel.%, but they also do not state that they are not on Telegram.
\end{itemize}

After the manual investigation, we mark as fakes or officials $96$ channels out of $502$.
In particular, among the $198$ channels recognized as fakes by our model, there are $24$ fakes, $161$ allegedly fakes, and $7$ official. While, among the channels classified as official, there are $57$ actual official channels, $239$ allegedly official, and $8$ fakes. Thus, for the channels we have evidence of their status, our classifier was able to classify $81$ channels out of $96$ correctly, equivalent to an accuracy of $84.3\%$, which aligned with the results obtained in cross-validation.

About the channels we verified to be fake, $17$ are of political figures from the Republican party, including $5$ claiming to be Donald Trump, $8$ are of celebrities, mostly actors, and $3$ are of news. 
Interestingly, $11$ fake channels, including those of actors and political figures, mainly forward messages about a conspiracy theory called Sabmyk. 

\section{SABMYK: a new conspiracy theory}
\label{sec:sabmyk}
Analyzing the fake channels detected by the model described in the previous section, we notice a group of $11$ channels related to \textit{Sabmyk}. This is a conspiracy theory that proposes itself as a better alternative to QAnon and promotes a singular quasi-religion centered around a messianic figure known as Sabmyk~\cite{independetSabmyk}. According to the \textit{"HOPE not hate"} organization, the Sabmyk-network has over a million members distributed on about one hundred Telegram channels~\cite{HopeSabmyk}. In particular, the mastermind of this operation is a German artist, Sebastian Bieniek, who has previously used social media to publicize his work.
Intrigued by the considerable number of members achieved by this conspiracy theory, we leverage the TGDataset to investigate how Sabmyk operated and how it built this strong network.

We start by discovering the Telegram channels involved in spreading the Sabmyk theory, leveraging the graph we built in Sec.~\ref{sec:graph_analysis} and a community detection algorithm.
A community in a graph is a subset of nodes that are densely connected to each other and weakly connected to nodes in other communities. To uncover the Sabmyk community, we used the Leiden algorithm~\cite{traag2019louvain}. 
At the end of the process, we find $91$ communities, of which one contains all the $11$ channels we already know.
The community discovered is made up of $98$ channels that, through manual investigation, we confirm are related to Sabmyk. Now, analyzing these channels, we examine how the people behind Sabmyk operate. 

\subsection{Spreading a Conspiracy Theory}
What is surprising about Sabmyk is how quickly it has spread. In a few months, these channels went from zero subscribers to an average of more than $10,000$, with the biggest channel \textit{Great Awakening Channel} with $133,815$. Concerning the temporal aspect, we find that the creation date of the first channel of the network is April 2020 and of the following $2$ channels is December 2020. However, it is only in 2021 that most of them appeared on Telegram ($64$ in January and $31$ in February).

\textbf{Attractive topics.}
The goal is to spread the messages as widely as possible. Thus, the need is to create channels dealing with topics that can attract many subscribers. As seen before, channels of services or public figures attract numerous subscribers, and it is challenging to distinguish an official channel from a fake or a clone one. Sabmyk exploited this idea by creating fake channels of famous people ($11\%$ of the network), institutional entities (\eg Department Of Defence, US Navy Channel, US Marines Channel), or news ($14\%$, \eg Liverpool Times, London Post, Chicago Reporter).
Another technique used was to create channels that target specific kinds of users near the Sabmyk theory. This category includes channels related to QAnon ($17\%$), far-right ($15\%$), or other conspiracies theories (\eg Obama Gate Truth, Chemtrails News).
In the Appendix (Tab.~\ref{tab:sabmyk_channels}), we report the complete list of Sabmyk-related channels we discovered.

\textbf{Reuse of content.} Producing content for about $100$ channels could be a laborious and time-consuming task.
In detail, $207,399$ messages have been shared within the Sabmyk network. However, analyzing these messages, we find that the number of distinct messages is only $5,038$. Indeed, most of the messages are forwarded multiple times within the network. The most shared messages are an image related to the "Great Awakening Channel" posted $8,622$ times ($4\%$ of total messages), the invitation link to join the channel of "John F. Kennedy Jr." posted $958$ times, and the message asking people to follow and share the Great Awakening Channel, posted $704$ times. 

\begin{figure}
  \subfigure[Message forwarding time]{%
  \includegraphics[width=.23\textwidth]{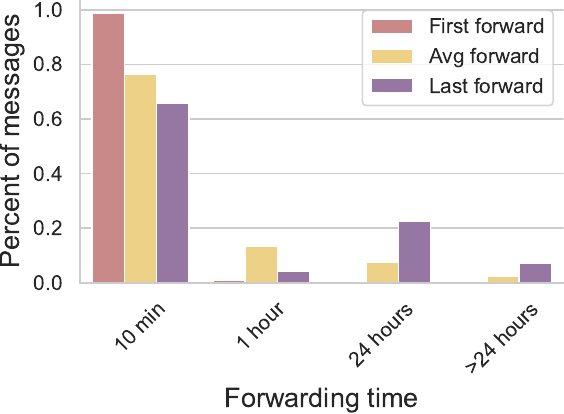}
  \label{fig:forwarding_pattern}}
  \subfigure[Copied messages]{%
  \includegraphics[width=.23\textwidth]{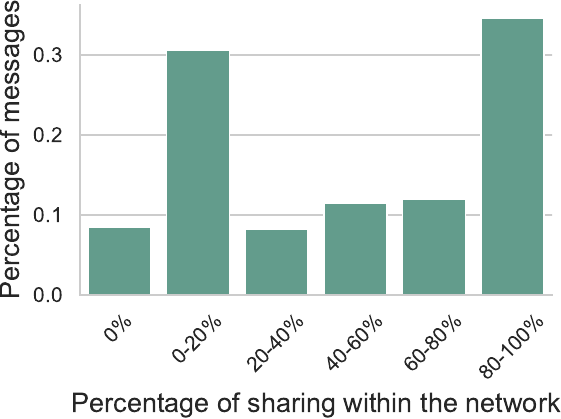}
  \label{fig:sabmyk_percentage_sharing}}
  \caption{Fig.~\ref{fig:forwarding_pattern} Forwarding time of first, average and last forwards of messages.~\ref{fig:sabmyk_percentage_sharing} Percentage of message sharing within the Sabmyk network.} 
  \label{fig:sabmyk}
\end{figure}

\textbf{(Almost) Random content.} 
Sabmyk intensively reuses its content. Through the analysis of forwarded messages of each channel, it is possible to note that there is no care about forwarding coherent messages with the channel's name. It is evident in channels with names of specific domains. For instance, in the channel named "Satoshi Nakamoto Official" (the pseudonym of the inventor of Bitcoin), $100\%$ of the forwarded messages are not related to the Bitcoin or cryptocurrency world. However, they pay attention to two aspects. The first point concerns the content created by the channel. Indeed, in this case, the content topic is related to the name of the channel. Looking again at the "Satoshi Nakamoto Official" channel, all the content created by this channel is related to the cryptocurrency world.
The second aspect is about the language of the messages. Indeed, the content written in English is forwarded over the whole network. Instead, messages written in other languages (\eg Italian, German) are forwarded only in the channels that specifically target users whose native language is the same as the content (\eg QAnonItaliano, Great Awakening DE).
Fig.~\ref{fig:sabmyk_percentage_sharing} shows the percentage of the network reached by each message. About $30\%$ of messages are shared between $20\%$ and $80\%$ of the network, while almost $34\%$ of messages by nearly the whole network.
Of particular interest are the messages that have never been forwarded ($0\%$ in the figure), accounting for about $8\%$. Indeed they are all messages that belong to the initial activity --- before the channels start to forward messages--- of the network's clone channels. Finally, the remaining $30\%$ of the messages forwarded by less than $20\%$ of the channels are the ones written not in English.

\textbf{High coordination.} The admins forward messages in the channels as soon as the new content is available. Fig.~\ref{fig:forwarding_pattern} shows the delay in forwarding messages from the creation time of the content. The first forward of a new message happens in the $98.6\%$ of the cases within $10$ minutes. It is likely because the content creator also manages other channels and instantaneously forwards the messages with them. 
The time that the whole network forwards a new message is incredibly fast. $65.8\%$ of messages cover the network in just $10$ minutes, and more than $90\%$ in the first 24 hours. Since the messages do not cover the whole network simultaneously, we believe that the forwarding is not managed by software or a single person but by many highly coordinated people.

%\begin{figure}
%  \includegraphics[width=.44\textwidth]{images/forwarding time barplot.pdf}
%  \caption{Forwarding time of first, average and last forwards of messages.}
%  \label{fig:forwarding_pattern}
%\end{figure}

\textbf{A core channel.} By analyzing the graph of the Sabmyk network as we did in Sec.~\ref{sec:graph_analysis}, we find that it consists of 2 strong connected components. One is of a single node, the channel entitled \textit{Sabmyk}, and the other component contains the remaining $97$ channels. Interestingly, the Sabmyk channel is the only one in the network that never forwards a message, whereas the whole network forwards all messages posted in the Sabmyk channel. Therefore, all the channels of the Sabmyk network are at 1 hop from the Sabmyk channel. Thus, it could be quite easy for users who joined one of the network channels to end up in the Sabmyk channel. Conversely, the users who joined the Sabmyk channel directly could remain unaware of the rest of the network.
Finally, observing the content created, we notice that, although every channel created at least one content, only $3$ channels created more than the $20\%$ of messages.

%\begin{figure}
%  \includegraphics[width=.44\textwidth]{images/bar_chart_coverage_forwarding.pdf}
%  \caption{Percentage of message sharing within the Sabmyk network.}
%  \label{fig:sabmyk_percentage_sharing}
%\end{figure}

%\section{Ethical Considerations}
%\label{sec:ethical_considerations}
%In this work, we analyzed $35,382$ channels on Telegram for a total number of more than $132$ million messages. The data collected were encrypted and stored for a limited amount of time in our servers. In addition, any reference to users, such as username or telephone number, was hashed to protect users' privacy further. Original data were not shared directly or placed on platforms that allow downloading. Consequently, and according to the policy of our IRB, we did not need any explicit authorization to perform our experiments.

\section{Related work}
There are several works focused on the Telegram ecosystem or emerging research issues related to it. Hashemi et al.~\cite{hashemi2019telegram} collect $900,000$  Iranian channels and $300,000$ Iranian groups on Telegram to identify high-quality groups, such as business groups, among low-quality groups (e.g., dating groups). They show that high-quality groups distinguish themselves from low-quality ones by longer messages and more user engagement. Nobari et al.~\cite{dargahi2017analysis} present a structural and topical analysis of messages posted on Telegram. In particular, they build a dataset of more than $2,000$ groups or channels and form a graph based on mentions. This study indicates that the PageRank algorithm is not suitable for detecting high-quality channels in Telegram. Jalilvand et al.~\cite{jalilvand2020channel}  address the problem of finding an ordered list of channels related to a user request in Telegram. 
%Ng et al.~\cite{ng2020pofma} analyze a Singapore-based COVID-19 Telegram group with more than $10,000$ participants and study the group opinion over time. Their work reveals that engagement reached a peak when the Ministry of Health increased the level of disease alert, but this involvement quickly declined.
Baumgartner et al.~\cite{baumgartner2020pushshift} publish a dataset of over $27,800$ thousand channels and $317,000$ messages from $2.2$ million unique users. Their dataset includes a wide range of right-wing extremist groups, as well as protest movements. In their work, Weerasinghe et al.~\cite{weerasinghe2020pod} reveal that Telegram hosts several organized groups, called pods, where each member interacts with each other's content to increase the popularity of their Instagram accounts. Other works~\cite{xu2019anatomy, la2020pump} reveal a vast presence on Telegram of channels and groups focused on cryptocurrency pump and dump. This market manipulation involves artificially inflating the price of a cryptocurrency held and then reselling it at a higher price.
Finally, several studies focus on the activity of terrorist organizations, like ISIS, that utilize Telegram for disseminating content and recruiting new followers~\cite{cao2017dynamical, yayla2017telegram}.

%\section{Limitations}
%The topics shown in Sec.~\ref{sec:topic_modeling} are not the main topics covered in general on Telegram. Instead, since we constructed the dataset by adding new channels based on the origin of the forwarded messages (Sec.~\ref{sec:bulding_dataset}), it suggests that messages of these topics are the ones most forwarded by other channels.

\section{Conclusions and Future works}
Telegram is becoming more popular every day, both as a classic instant messaging app and as a platform to deliver live updates and content to a large audience. Thus, it becomes increasingly important to understand what happens on the platform and how it will evolve in the future.

In this paper, we present our dataset of Telegram's channels, the TGDataset, made of more than $35,000$ channels. To the best of our knowledge, this is the first dataset of Telegram's channels that attempt to make a general-purpose snapshot of the channels present on Telegram. Thus we will release it publicly accessible~\cite{dataset} to help researchers on further investigations. 
Starting to scratch the surface of our dataset, we discovered the presence of several borderline activities running their business on public channels, as well as channels dealing with dubious ethical content. 
In our study, we mainly focused on the widespread phenomenon of fake and clone channels. We characterize these kinds of channels and try to understand how admins of these channels take advantage of them. 
We propose a machine learning model that achieves an accuracy of $86\%$ in detecting fake accounts. Running our detector on a subset of our dataset, we found $191$ allegedly fake account of which we could confirm $27$.
Given the extent of the phenomenon and the difficulty of distinguishing fake channels from official ones, even for technically knowledgeable users, the need for institutions, famous people, and organizations to obtain the verified status for their channels is on the rise. Indeed, we notice assessing our results that only few official channels leverage this opportunity.
Finally, we investigated Sabmyk, a conspiracy theory that exploited fake and clone channels to reach a broad audience.

With this work, we shed light on several dark sides of Telegram. However, we believe further investigations are needed to illuminate the Telegram ecosystem completely. Indeed, in our research, we leverage only text-based messages. Considering media-based content, we think it is possible to achieve a more refined topic modeling classification. Moreover, we believe Telegram public groups are a vast portion of Telegram and deserve further exploration. Indeed, here it is possible to easily access the complete list of subscribers, compromising the users' privacy and impersonating the administrators to carry out frauds.

\printbibliography

@article{yayla2017telegram,
  title={Telegram: The mighty application that ISIS loves},
  author={Yayla, Ahmet S and Speckhard, Anne},
  journal={International Center for the Study of Violent Extremism},
  year={2017}
}

@inproceedings{mimno2011optimizing,
  title={Optimizing semantic coherence in topic models},
  author={Mimno, David and Wallach, Hanna and Talley, Edmund and Leenders, Miriam and McCallum, Andrew},
  booktitle={Proceedings of the 2011 conference on empirical methods in natural language processing},
  pages={262--272},
  year={2011}
}

@article{gardner1998artificial,
  title={Artificial neural networks (the multilayer perceptron)—a review of applications in the atmospheric sciences},
  author={Gardner, Matt W and Dorling, SR},
  journal={Atmospheric environment},
  volume={32},
  number={14-15},
  pages={2627--2636},
  year={1998},
  publisher={Elsevier}
}

@inproceedings{mannor2005cross,
  title={The cross entropy method for classification},
  author={Mannor, Shie and Peleg, Dori and Rubinstein, Reuven},
  booktitle={Proceedings of the 22nd international conference on Machine learning},
  pages={561--568},
  year={2005}
}

@inproceedings{zhang2018improved,
  title={Improved adam optimizer for deep neural networks},
  author={Zhang, Zijun},
  booktitle={2018 IEEE/ACM 26th International Symposium on Quality of Service (IWQoS)},
  pages={1--2},
  year={2018},
  organization={IEEE}
}

@inproceedings{hara2015analysis,
  title={Analysis of function of rectified linear unit used in deep learning},
  author={Hara, Kazuyuki and Saito, Daisuke and Shouno, Hayaru},
  booktitle={2015 international joint conference on neural networks (IJCNN)},
  pages={1--8},
  year={2015},
  organization={IEEE}
}

@inproceedings{cao2012aiding,
  title={Aiding the detection of fake accounts in large scale social online services},
  author={Cao, Qiang and Sirivianos, Michael and Yang, Xiaowei and Pregueiro, Tiago},
  booktitle={9th $\{$USENIX$\}$ Symposium on Networked Systems Design and Implementation ($\{$NSDI$\}$ 12)},
  pages={197--210},
  year={2012}
}

@inproceedings{xiao2015detecting,
  title={Detecting clusters of fake accounts in online social networks},
  author={Xiao, Cao and Freeman, David Mandell and Hwa, Theodore},
  booktitle={Proceedings of the 8th ACM Workshop on Artificial Intelligence and Security},
  pages={91--101},
  year={2015}
}

@misc{nakatani2010langdetect, title = {Language Detection Library for Java}, author = {Shuyo, Nakatani}, url = {http://code.google.com/p/language-detection/}, year = {2010} }

@misc{dataset, title = {Telegram Dataset}, author = {Anonymous}, url = {}, year = {2021} }

@article{blei2003latent,
  title={Latent dirichlet allocation},
  author={Blei, David M and Ng, Andrew Y and Jordan, Michael I},
  journal={the Journal of machine Learning research},
  volume={3},
  pages={993--1022},
  year={2003},
  publisher={JMLR. org}
}

@article{kigerl2020behind,
  title={Behind the Scenes of the Underworld: Hierarchical Clustering of Two Leaked Carding Forum Databases},
  author={Kigerl, Alex},
  journal={Social Science Computer Review},
  pages={0894439320924735},
  year={2020},
  publisher={SAGE Publications Sage CA: Los Angeles, CA}
}

@inproceedings{stevens2012exploring,
  title={Exploring topic coherence over many models and many topics},
  author={Stevens, Keith and Kegelmeyer, Philip and Andrzejewski, David and Buttler, David},
  booktitle={Proceedings of the 2012 joint conference on empirical methods in natural language processing and computational natural language learning},
  pages={952--961},
  year={2012}
}

@article{hofmann2001unsupervised,
  title={Unsupervised learning by probabilistic latent semantic analysis},
  author={Hofmann, Thomas},
  journal={Machine learning},
  volume={42},
  number={1},
  pages={177--196},
  year={2001},
  publisher={Springer}
}

@inproceedings{lundberg2017unified,
  title={A unified approach to interpreting model predictions},
  author={Lundberg, Scott M and Lee, Su-In},
  booktitle={Proceedings of the 31st international conference on neural information processing systems},
  pages={4768--4777},
  year={2017}
}

@misc{TelegramUsers,
  title={Telegram FAQ},
  howpublished = {\url{https://telegram.org/faq-what-is-telegram-what-do-i-do-here}},
  year={2021}
}

@misc{telemeterIO,
  title={TelemeterIo},
  howpublished = {\url{https://telemetr.io/en/channels}},
  year={2021}
}

@misc{tgstat,
  title={Telegram Analytics},
  howpublished = {\url{https://tgstat.com/}},
  year={2021}
}

@misc{telethonAPI,
  title={Telethon’s Documentation},
  howpublished = {\url{https://docs.telethon.dev/en/latest/}}
}

@misc{howVeryfyTelegramCh,
  title={Page Verification Guidelines},
  howpublished = {\url{https://telegram.org/verify}}
}

@misc{telegramNeoNazi,
  title = {Telegram the latest safe haven for white supremacists},
  howpublished = {\url{https://www.adl.org/blog/telegram-the-latest-safe-haven-for-white-supremacists}},
  year={2019}
}

@misc{fortune,
  title = {WhatsApp delays privacy policy changes after users defect to rivals Signal and Telegram},
  howpublished = {\url{https://fortune.com/2021/01/15/whatsapp-delays-privacy-policy-changes-after-users-defect-to-rivals-signal-and-telegram/}}
}

@misc{bbcterrorist,
  title = {Telegram to block terror channels after Indonesian ban},
  howpublished = {\url{https://www.bbc.com/news/business-40627739}}
}

@misc{revengePorn,
  title = {Telegram's massive revenge porn problem has made these women’s lives hell},
  howpublished = {\url{https://mashable.com/article/nudes-revenge-porn-crime-telegram}},
  year={2020}
}

@misc{chlidporn,
  title = {Leaked sex tapes and child porn: A look into 13 illicit Telegram chat groups},
  howpublished = {\url{https://www.channelnewsasia.com/singapore/telegram-chat-groups-nasi-lemak-porn-upskirt-sex-police-846911}},
  year={2019}
}

@misc{telegramScam,
  title = {Scammers in telegram and how to report},
  howpublished = {\url{https://www.telegramadviser.com/scammers-in-telegram-and-how-to-report/}}
}

@misc{covidTelegram,
  title = {Telegram, the powerful COVID-19 choice of communications by many governments},
  howpublished = {\url{https://www.channelnewsasia.com/commentary/coronavirus-covid-19-government-telegram-whatsapp-fake-news-info-936061}}
}

@misc{HopeSabmyk,
  title = {Unmasked: the QAnon ‘messiah’},
  howpublished = {\url{https://www.hopenothate.org.uk/unmasked-the-qanon-messiah//}}
}

@misc{independetSabmyk,
  title = {Sabmyk Network: Founder of bizarre new religion targeting QAnon believers ‘unmasked’ by Hope Not Hate},
  howpublished = {\url{https://www.independent.co.uk/news/world/europe/sabmyk-network-qanon-conspiracy-theories-b1820639.html}}
}

@article{hartigan1979ak,
  title={AK-means clustering algorithm},
  author={Hartigan, John A and Wong, Manchek A},
  journal={Journal of the Royal Statistical Society: Series C (Applied Statistics)},
  volume={28},
  number={1},
  pages={100--108},
  year={1979},
  publisher={Wiley Online Library}
}

@inproceedings{kempe2005influential,
  title={Influential nodes in a diffusion model for social networks},
  author={Kempe, David and Kleinberg, Jon and Tardos, {\'E}va},
  booktitle={International Colloquium on Automata, Languages, and Programming},
  pages={1127--1138},
  year={2005},
  organization={Springer}
}

@article{chen2013identifying,
  title={Identifying influential nodes in large-scale directed networks: the role of clustering},
  author={Chen, Duan-Bing and Gao, Hui and L{\"u}, Linyuan and Zhou, Tao},
  journal={PloS one},
  volume={8},
  number={10},
  pages={e77455},
  year={2013},
  publisher={Public Library of Science}
}

@article{traag2019louvain,
  title={From Louvain to Leiden: guaranteeing well-connected communities},
  author={Traag, Vincent A and Waltman, Ludo and Van Eck, Nees Jan},
  journal={Scientific reports},
  volume={9},
  number={1},
  pages={1--12},
  year={2019},
  publisher={Nature Publishing Group}
}

@article{brin1998anatomy,
  title={The anatomy of a large-scale hypertextual web search engine},
  author={Brin, Sergey and Page, Lawrence},
  journal={Computer networks and ISDN systems},
  volume={30},
  number={1-7},
  pages={107--117},
  year={1998},
  publisher={Elsevier}
}

@article{hashemi2019telegram,
  title={Telegram group quality measurement by user behavior analysis},
  author={Hashemi, Ali and Chahooki, Mohammad Ali Zare},
  journal={Social Network Analysis and Mining},
  volume={9},
  number={1},
  pages={1--12},
  year={2019},
  publisher={Springer}
}

@inproceedings{dargahi2017analysis,
  title={Analysis of Telegram, an instant messaging service},
  author={Dargahi Nobari, Arash and Reshadatmand, Negar and Neshati, Mahmood},
  booktitle={Proceedings of the 2017 ACM on Conference on Information and Knowledge Management},
  pages={2035--2038},
  year={2017}
}

@inproceedings{baumgartner2020pushshift,
  title={The Pushshift Telegram Dataset},
  author={Baumgartner, Jason and Zannettou, Savvas and Squire, Megan and Blackburn, Jeremy},
  booktitle={Proceedings of the International AAAI Conference on Web and Social Media},
  volume={14},
  pages={840--847},
  year={2020}
}

@inproceedings{weerasinghe2020pod,
  title={The pod people: Understanding manipulation of social media popularity via reciprocity abuse},
  author={Weerasinghe, Janith and Flanigan, Bailey and Stein, Aviel and McCoy, Damon and Greenstadt, Rachel},
  booktitle={Proceedings of The Web Conference 2020},
  pages={1874--1884},
  year={2020}
}

@article{cao2017dynamical,
  title={Dynamical patterns in individual trajectories toward extremism},
  author={Cao, Zhenfeng and Zheng, Minzhang and Vorobyeva, Yulia and Song, Chaoming and Johnson, Neil},
  journal={Available at SSRN 2979345},
  year={2017}
}

@article{hoseini2021globalization,
  title={On the Globalization of the QAnon Conspiracy Theory Through Telegram},
  author={Hoseini, Mohamad and Melo, Philipe and Benevenuto, Fabricio and Feldmann, Anja and Zannettou, Savvas},
  journal={arXiv preprint arXiv:2105.13020},
  year={2021}
}

@article{jalilvand2020channel,
  title={Channel retrieval: finding relevant broadcasters on Telegram},
  author={Jalilvand, Asal and Neshati, Mahmood},
  journal={Social Network Analysis and Mining},
  volume={10},
  number={1},
  pages={1--16},
  year={2020},
  publisher={Springer}
}

@inproceedings{la2020pump,
  title={Pump and Dumps in the Bitcoin Era: Real Time Detection of Cryptocurrency Market Manipulations},
  author={La Morgia, Massimo and Mei, Alessandro and Sassi, Francesco and Stefa, Julinda},
  booktitle={2020 29th International Conference on Computer Communications and Networks (ICCCN)},
  pages={1--9},
  year={2020},
  organization={IEEE}
}

@inproceedings{xu2019anatomy,
  title={The anatomy of a cryptocurrency pump-and-dump scheme},
  author={Xu, Jiahua and Livshits, Benjamin},
  booktitle={28th $\{$USENIX$\}$ Security Symposium ($\{$USENIX$\}$ Security 19)},
  pages={1609--1625},
  year={2019}
}

@inproceedings{anguita2012k,
  title={The ‘K’in K-fold cross validation},
  author={Anguita, Davide and Ghelardoni, Luca and Ghio, Alessandro and Oneto, Luca and Ridella, Sandro},
  booktitle={20th European Symposium on Artificial Neural Networks, Computational Intelligence and Machine Learning (ESANN)},
  pages={441--446},
  year={2012},
  organization={i6doc. com publ}
}

% or

\clearpage
\appendix %[Appendix]

\onecolumn
\section{The Sabmyk network}

\begin{table*}[h]
    \centering
    
\pgfplotstabletypeset[
    every head row/.style={before row=\toprule,after row=\midrule},
    every last row/.style={after row=\bottomrule},
    col sep=&,
    row sep=\\,
    columns={Title,Subscribers,Username,Title,Subscribers,Username},
    display columns/0/.style={
        % first part of 2 of `thing':
        select equal part entry of={0}{2},
        string type,
            % column display name:
        column name={Title},
        column type={l}, % ... and type
    },
    display columns/1/.style={
        % first part of 2 of `mapsto':
        select equal part entry of={0}{2},
        numeric type,
        column name={Subscribers},
        column type={c},
    },
    display columns/2/.style={select equal part entry of={0}{2},string type,column name={Username},column type={c|},},
    display columns/3/.style={select equal part entry of={1}{2},string type, column name={Title}, column type={l}},% second part of 2 of `thing'
    display columns/4/.style={select equal part entry of={1}{2},numeric type, column name={Subscribers}},% second part of 2 of `maps'
    display columns/5/.style={select equal part entry of={1}{2},string type, column name={Username},},% second part of 2 of `maps'
]{
    Title&Subscribers&Username\\
American Tribune&3733&AmericanTribune\\
Anti Corona Terrorism&6485&AntiCoronaTerrorism\\
Anti Corona-Regime&6184&AntiCoronaRegime\\
Anti Fake Pandemic&4345&AntiFakePandemic\\
AntiGates&14929&AntiGates\\
Antiilluminati Official&11524&AntiilluminatiOfficial\\
Art is all around&973&artisallaround\\
Atlantis Official&882&AtlantisOfficial\\
Atmumra&3371&atmumra\\
Awaken We Are&2697&AwakenWeAre\\
Bravetower&1314&Bravetower\\
British Patriots Party&9408&BritishPatriotsParty\\
Capitol News&7410&CapitolNews\\
Charles Flynn&4403&CharlesFlynn\\
Chemtrails News&519&ChemtrailsNews\\
Chicago Reporter&2607&ChicagoReporter\\
Clint Eastwood Real&2638&ClintEastwoodReal\\
Das Grosse E&745&DasGrosseE\\
Department of Defence&5478&DepartmentOfDefence\\
Digital Supersoldier&5353&DigitalSupersoldier\\
Donald J. Trump Team&14923&DTrumpt\\
Donald Trump Jr.&15979&DonaldTrumpJrInfo\\
Drain The Swamp News&20287&DrainTheSwampNews\\
Electric News Channel&1433&ElectricNewsChannel\\
England First&290&EnglandFirst\\
First Flush News&5110&FirstFlushNews\\
General Flynn Info&16856&GeneralFlynnInfo\\
\textbf{Great Awakening Channel} & 133815&GreatAwakeningChannel\\
Great Awakening DE&2849&GreatAwakeningDe\\
Great Awakening France&255&GreatAwakeningFrance\\
Great Awakening Official&18904&GreatAwakeningOfficial\\
Great Awakening UK&1070&GreatAwakeningUK\\
Great Awakening US&2617&GreatAwakeningUS\\
Great Corona Coup&2955&GreatCoronaCoup\\
Hardhauer&1995&Hardhauer\\
J. C. Miller&6084&JCMiller\\
Joe M&25515&HereJoeM\\
John F. Kennedy Jr.&76097&JohnFKennedyJr\\
Jon Voight Real&3238&JonVoightReal\\
Josh Hawley&867&RealJoshHawley\\
Kanye West&204&KanyeOW\\
Keanu Reeves Real&10801&KeanuReevesReal\\
L. Lin Wood&15850&LLWoodChannel\\
Liberty Online News&1934&LibertyOnlineNews\\
Liverpool Times&1738&LiverpoolTimes\\
London Post&1659&LondonPost\\
Los Angeles Post&2065&LosAngelesPost\\
MAGA&7054&GreaterMAGA\\
Mel Gibson Real&13822&MelGibsonReal\\
Memepow&3939&memepow\\
Michael Jackson Alive&5442&MichaelJacksonAlive\\
Mike Lindell&14406&MikeJLindell\\
Nesara Gesara Info&13714&NesaraGesaraInfo\\
Nicola Tesla News&12779&NicolaTeslaNews\\
Noah's Prophecy&20224&NoahsProphecy\\
Nostradamus Info&2323&NostradamusInfo\\
ObamaGate Truth&4334&ObamaGateTruth\\
Official Anonymous&6634&OfficialAnonymous\\
Official Times&6882&OfficialTimes\\
Patriot Party US&11756&PatriotPartyUS\\
Q&4382&HereIsQ\\
Q Donald J. Trump&16739&QDonaldJTrump\\
Q Drops Feed&10239&QdropsFeed\\
Q speaking&16342&qspeaking\\
QAnon Central&16526&QAnonCentral\\
QAnon DEU&2824&QAnonDEU\\
QAnon People&6617&QAnonPeople\\
QAnon Storm Base&27667&QAnonStormBase\\
QAnon italiano&670&QAnonItaliano\\
Quotation Nation&753&quotationnation\\
Ron DeSantis&5272&RealRonDeSantis\\
Rudy Giuliani&57494&RGiuliani\\
Sabmyk&21123&sabmyk\\
Sabmyk Awakening&1342&SabmykAwakening\\
Satoshi Nakamoto Official&4396&OfficialSatoshi\\
Secretary Mike Pompeo&11258&RealMikePompeo\\
Shawunawaz&4517&shawunawaz\\
Sidney Powell&23014&SideyPowellAcount\\
Space Force News&13039&SpaceForceNews\\
Starseed Children&4653&StarseedChildren\\
Steve Bannon&3199&RealSteveBannon\\
Supernarrativ&2109&supernarrativ\\
Sylvester Stallone&1346&SylvesterSt\\
T. O'Sullivan&3490&TOSullivan\\
The Great Awakening&14045&TheGreatAwakening2\\
The Real Awakening&12082&TheRealAwakening\\
The Trumpists&12347&TheTrumpists\\
Tom Brady Real&1305&TomBradyReal\\
Tucker Carlson News&1503&TuckerCarlsonNews\\
U. S. Marines Channel&4650&USmarinesChannel\\
U. S. Navy Channel&2125&USnavyChannel\\
US Military Voice&29889&USmilitaryVoice\\
US Patriots&16437&USPatriots\\
Ufo Official&1598&UfoOfficial\\
WWG1WGA here&19485&WWG1WGAhere\\
We Are The Faithful&9220&WeAreTheFaithful\\
Wiki Official&4987&WikiOfficial\\
YellowStoneWolf&3667&YellowStoneWolf\\
}
    \caption{Titles, number of subscribers, and username of the Sabmyk channels. The largest channel of the network is in bold. Prepending the string \textit{https://t.me/} to the username is possible to obtain the URL of the channel (https://t.me/username).}
    \label{tab:sabmyk_channels}
\end{table*}

\section{Topic discovered within our dataset}

%\section{Topic discovered within our dataset}
\begin{table*}[h]
    \centering

    \begin{tabular}{l|c}
    \toprule
        Topic & \multicolumn{0}{c}{Top 10 keywords} \\
        \midrule
        Porn      & videos, interview, drop, porn, fetish, queen, teen, click, vent, bitch   \\ 
        Religion and supremacist & 4chan, jew, jesus, christ, states, hitler, church, lord, proud, identity, antifa    \\
        Software &   php, linux, web, vulnerability, item, readhacker, scan, software, privacy, hours \\
        Carding &  trick, iphone, apple, carding, amazon, method, samsung, payment, cracking, prime  \\
        Telephone modding & xda, feedproxy, appeared, android, galaxy, samsung, developer, download, pixel, oneplus   \\
        Hacking & premium, proxy, mod, click, hack, server, netflix, port, log, apk   \\
        Cryptocurrencies &  reddit, bitcoin, btc, ift, crypto, rss, eth, xrp, stock, exchange  \\
        Games hacking &  root, pubg, password, hack, mod, aimbot, antiban, esp, mobile, recoil  \\
        India news/career & india, upsc, minister, affairs, foreign, exam, promotion, articleshow, express, hindu   \\
        Trump supporters &  tweet, donald, realdonaldtrump, joebiden, zerohedge, hashtag, tribunal, dbongino, wire, february, shapiro  \\
        News &   biden, coronavirus, vaccine, reuters, election, donald, court, pandemic, capitol, lockdown \\
        Free music/movie &  download, 720p, genre, artist, audio, tumblr, title, album, size, spotify  \\
        Violated terms/pornographic &  displayed, violated, terms, pornographic, post, placed, rockwell, infringement, unavailable, copyright  \\
        Games discussion &  castle, defense, gold, lvl, protected, points, fox, successfully, potato, pouch  \\
        \bottomrule
    \end{tabular}
    \caption{Top 10 keywords within LDA topics.}
    \label{tab:LDA_topics}
\end{table*}

%\section{Topic discovered within our dataset}

\end{document}